\documentclass[review]{elsarticle}

\usepackage{lineno,hyperref}
\usepackage[nolist]{acronym}
\usepackage[onehalfspacing]{setspace}
\usepackage{booktabs}
\usepackage{float}
\usepackage{amsmath}
\usepackage{listings}
\modulolinenumbers[1]

\journal{Astroparticle Physics}

\begin{document}

\begin{frontmatter}

\title{Normalized and Asynchronous Mirror Alignment\\ for Cherenkov Telescopes}
\author[a]{M.~L.~Ahnen}
\author[d]{D.~Baack}
\author[b]{M.~Balbo}
\author[c]{M.~Bergmann}
\author[a]{A.~Biland}
\author[c]{M.~Blank}
\author[a]{T.~Bretz}
\author[d]{K.~A.~Bruegge}
\author[d]{J.~Buss}
\author[d]{M.~Domke}
\author[c]{D.~Dorner}
\author[d]{S.~Einecke}
\author[c]{C.~Hempfling}
\author[a]{D.~Hildebrand}
\author[a]{G.~Hughes}
\author[a]{W.~Lustermann}
\author[c]{K.~Mannheim}
\author[a]{S.~A.~Mueller\corref{cor1}}
\author[a]{D.~Neise}
\author[b]{A.~Neronov}
\author[d]{M.~Noethe}
\author[d]{A.-K.~Overkemping}
\author[c]{A.~Paravac}
\author[a]{F.~Pauss}
\author[d]{W.~Rhode}
\author[a]{A.~Shukla}
\author[d]{F.~Temme}
\author[d]{J.~Thaele}
\author[b]{S.~Toscano}
\author[a]{P.~Vogler}
\author[b]{R.~Walter}
\author[c]{A.~Wilbert}
\cortext[mycorrespondingauthor]{S. A. Mueller\\sebmuell@phys.ethz.ch}
\address[a]{ETH Zurich, Institute for Particle Physics\\
Otto-Stern-Weg 5, 8093 Zurich, Switzerland}
\address[b]{University of Geneva,  ISDC Data Center for Astrophysics\\
 Chemin d'Ecogia 16,  1290 Versoix,  Switzerland}
\address[c]{Universit\"at W\"urzburg, Institute for Theoretical Physics and Astrophysics\\
Emil-Fischer-Str. 31, 97074 W\"urzburg,  Germany}
\address[d]{TU Dortmund, Experimental Physics 5\\
Otto-Hahn-Str. 4, 44221 Dortmund, Germany}
%
\begin{abstract}
\acfp{iact} need imaging optics with large apertures and high image intensities
to map the faint Cherenkov light emitted from cosmic ray air showers onto their
image sensors. 
Segmented reflectors fulfill these needs, and as they are composed from mass production mirror facets they are inexpensive and lightweight. 
However, as the overall image is a superposition of the individual facet images, alignment is a challenge. 
Here we present a computer vision based star tracking alignment method, which also works for limited or changing star light visibility.
Our method normalizes the mirror facet reflection intensities to become independent of the reference star's intensity or the cloud coverage.
Using two CCD cameras, our method records the mirror facet orientations asynchronously of the telescope drive system, and thus makes the method easy to integrate into existing telescopes.
It can be combined with remote facet actuation, but does not require one to work.
Furthermore, it can reconstruct all individual mirror facet point spread functions without moving any mirror.
We present alignment results on the $4\,$meter \acf{fact}.
\end{abstract}
\begin{keyword}
Mirror alignment, computer vision, point spread function
\end{keyword}
\end{frontmatter}
\newcommand{\RadioCoeffCresponse}{r_\text{resp}}
\newcommand{\RadioCoeffTexp}{r_\text{expo}}
\newcommand{\RadioConstant}{r_\text{const}}
\newcommand{\NormalizedMirrorResponse}{R}
\newcommand{\StarIntensity}{s}
\newcommand{\MirrorReflectionIntensity}{m}
\newcommand{\RelativePointing}{\Theta}
\newcommand{\texp}{T_\text{expo}}
\newcommand{\iflux}{I_\text{pix}}
\newcommand{\cresponse}{C_\text{pix}}
\newcommand{\geom}{\alpha}
\newcommand{\FigCapLabSca}[4]{
    \begin{figure}[H]
        \begin{center}
            \includegraphics[width=#4\textwidth]{#1}
            \caption[]{#2}
            \label{#3}
        \end{center}
    \end{figure}
}
\newcommand{\FigCapLab}[3]{
    \FigCapLabSca{#1}{#2}{#3}{1.0}
}
\newcommand{\TwoFigsSideBySide}[2]{
    \begin{minipage}[t]{0.485\linewidth}
        \vspace{-0.5cm}
        \includegraphics[width=1\textwidth]{#1}
    \end{minipage}
    \hfill
    \begin{minipage}[t]{0.485\linewidth}
        \vspace{-0.5cm}
        \includegraphics[width=1\textwidth]{#2}
        \vspace{-1cm}
    \end{minipage}
}
\newcommand{\SideBySide}[2]{
    \newline
    \begin{minipage}[t]{0.485\linewidth}
        #1
        \end{minipage}
    \hfill
        \begin{minipage}[t]{0.485\linewidth}
        #2
    \end{minipage}\\
}
\section{Introduction}
The \acf{iact} technique, with its large effective area, has opened the very high energy gamma ray sky to astronomy.
Almost \cite{CANGAROO1_optics} all former \cite{WHIPPLE_optics, CAT_Themis_optics}, current \cite{VERITAS_optics, HESS_I_optics, HESS_II_optics, MAGIC_optics, FACT_design}, and future \acsp{iact} \cite{CTA_Introduction, TAIGA_IACT_optics} make use of segmented imaging reflectors with apertures up to $614\,$m$^2$ \cite{, HESS_II_optics}.
Segmented reflector facets can be mass produced cost-efficiently with an acceptable image quality.
Facets are much lighter than a monolithic mirror, which allows for very fast telescope repositioning, e.g. for gamma ray burst hunting.\\
However, there is one challenge to segmented reflectors.
This is the task of manipulating the mirror facet orientations and positions to improve the spatial image resolution or to reduce the \ac{psf}'s size, also known as alignment.
Alignment needs to be done not only during installation but also in case of repair and replacement of facets.
To find the few gamma ray induced events in the far more numerous class of hadronic cosmic ray induced events, the air shower records are analyzed for geometrical and temporal features.
Higher image contrast, more isochronous photon arrivals, less image distortions, and higher spatial image resolution all help to lower the trigger and energy threshold of an \acs{iact} and further help to better reconstruct the energies and source positions of the primary particles.
This makes alignment and mirror facet orientation determination important for an \acs{iact}.\\
Above this, determining the mirror facet orientations without manipulation of the facets orientations or positions is desirable to put the mirror facet orientations in the \acs{iact} simulations, which rely on precise input to calibrate the \acs{iact}.
%
\section{Current methods}
\label{secCurrentAlignmentMethods}
To tackle the challenge of alignment, several approaches are in use.
We can summarize them in three categories.
\newline
First, there is the 2$f$ method and its more general equivalent Bokeh alignment \cite{FACT_Bokeh_alignment}.
Both can be performed with simple hardware and minimal to no software.  
With an artificial light source, the mirror facets are aligned without the need for clear nights or star light.
However, 2$f$ alignment is geometrically very restricted \cite{VERITAS_bias_alignment, CAT_Themis_optics} and struggles to reach the most inner facets. 
The novel Bokeh alignment is less restricted but was not yet performed on typical observation elevations to correct for gravitational slump.
In the first simple implementation of Bokeh alignment on an \acs{iact}, which was limited by the thin lens approximation, the method did not yet reach the same high quality \acs{psf} as the star tracking methods \cite{FACT_Bokeh_alignment}.
\newline
Second, there are methods which do direct \acs{psf} investigations on a tracked star using a dedicated image screen \cite{MAGIC_amc, HESS_II_optics}.
While achieving high \acs{psf} quality these methods make extensive use of mirror facet orientation actuation during alignment.
Because of overcoming the \acs{psf} facet ambiguity using the mirror facet actuation, the run time of these methods scales linear with the number of facets.
\newline
Third, there is the \acs{veritas} raster method \cite{VERITAS_SCCAN_alignment} which is based on the \acs{sccan} \cite{SCCAN_Arqueros_2005} method.
Synchronized with the telescope's drive, the telescope rasters a grid on the sky dome close to a star while the mirror facets surfaces are recorded by a CCD camera, positioned in the reflector's focal point.
In contrast to the direct \acs{psf} observations, many different pointing positions are needed to restore the mirror facet \acsp{psf} but on the other hand the raster method's run time is constant in the number of facets.
Since the facet reflections are recorded through out the whole process of grid pointing, the method works best when the reference star's light intensity is steady over time, which favors clear nights and sticking to one particular reference star. 
Combining two records on different reference stars is not straight forward since the stars might have different light intensities.
Also there is extensive bi-directional communication to the telescope drive, which needs a software interface. 
\newline
Here we present an enhanced alignment system based on the \acs{veritas} raster scan method.
We implement a second CCD camera, that observes the reference star.
Using the star's intensity observed in the second camera, our method normalizes the facet reflection intensities.
Further the second camera is used to estimate the telescope's relative pointing, which makes communication to the telescope drive obsolete by recording asynchronously of it.
We call it \acf{namod}.
\acs{namod} can combine records taken on different reference stars flexibly.
\acs{namod} can reconstruct all individual mirror facet \acsp{psf} and facet orientations without needing actuated facets.
It delivers high \acs{psf} quality while allowing alignment for a wider range of night sky conditions.
We present our implementation of \acs{namod} on the 10\,th scale mock up Mini \acs{fact} and our \acs{namod} implementation's alignment performance on the \acf{fact}, a $4\,$m class \ac{iact} on the Canary island La Palma, Spain.
%
\section{Method and Implementation}
\label{SecApparatus}
Our \acs{namod} implementation uses two digital cameras, see Figure \ref{FcamOnRefl}. 
First, the reflector camera monitors all the reflector's facets and is mounted in the focal point of the reflector.
Second, the star camera is mounted parallel to the telescope's optical axis and observes the same part of the night sky as the telescope.
\FigCapLabSca{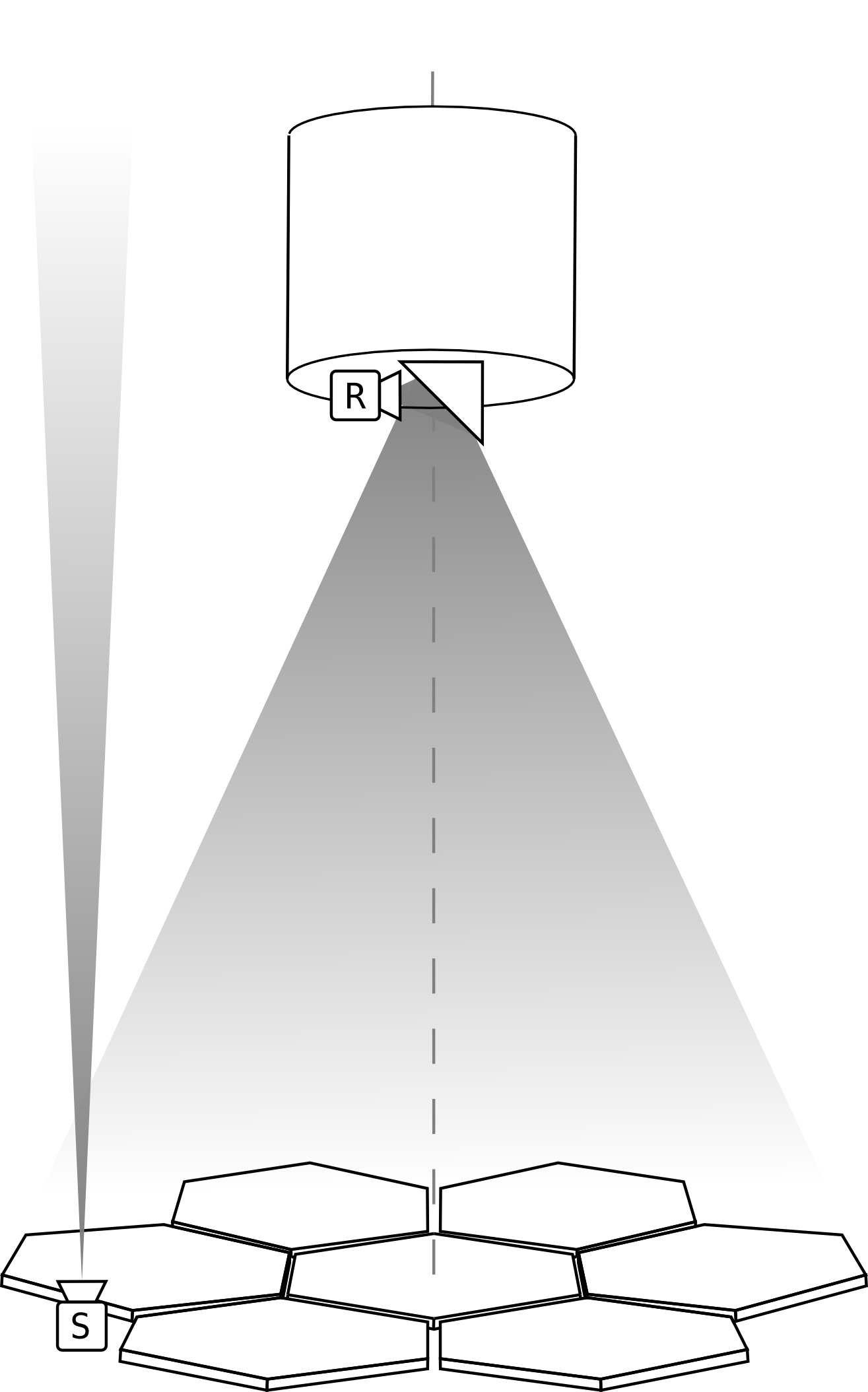}{%
    \acs{namod}'s star camera [S] and reflector camera [R] mounted on a telescope with a segmented reflector. 
    The reflector camera is mounted in the pseudo focal point provided by a 45$^\circ$ mirror. 
    In this way the telescope's image sensor can stay in place.
    The shaded triangles represent the cameras view cones.
}{FcamOnRefl}{0.6}
\FigCapLab{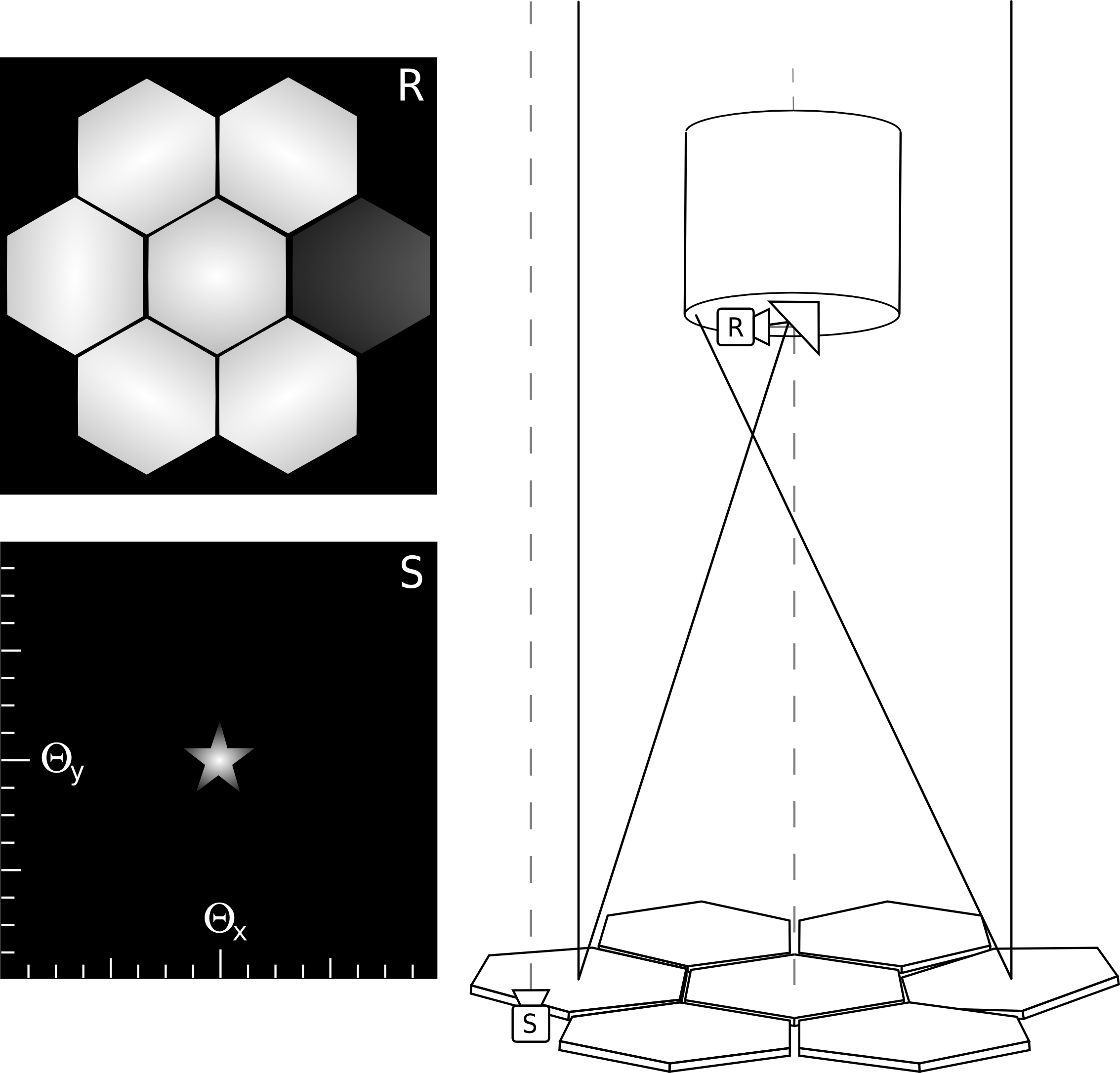}{%
    Star light rays and \acs{namod} images of star [S] and reflector [R] camera when pointing on the star. 
    The star appears in the center of the star camera image.
    The mirror facet on the right is misaligned. 
    Its reflected light does not reach the reflector's focal point, and it appears dark in the reflector image. 
    All the other facets are aligned well and therefore look bright.
}{FonAxisNamodSchematic}
\FigCapLab{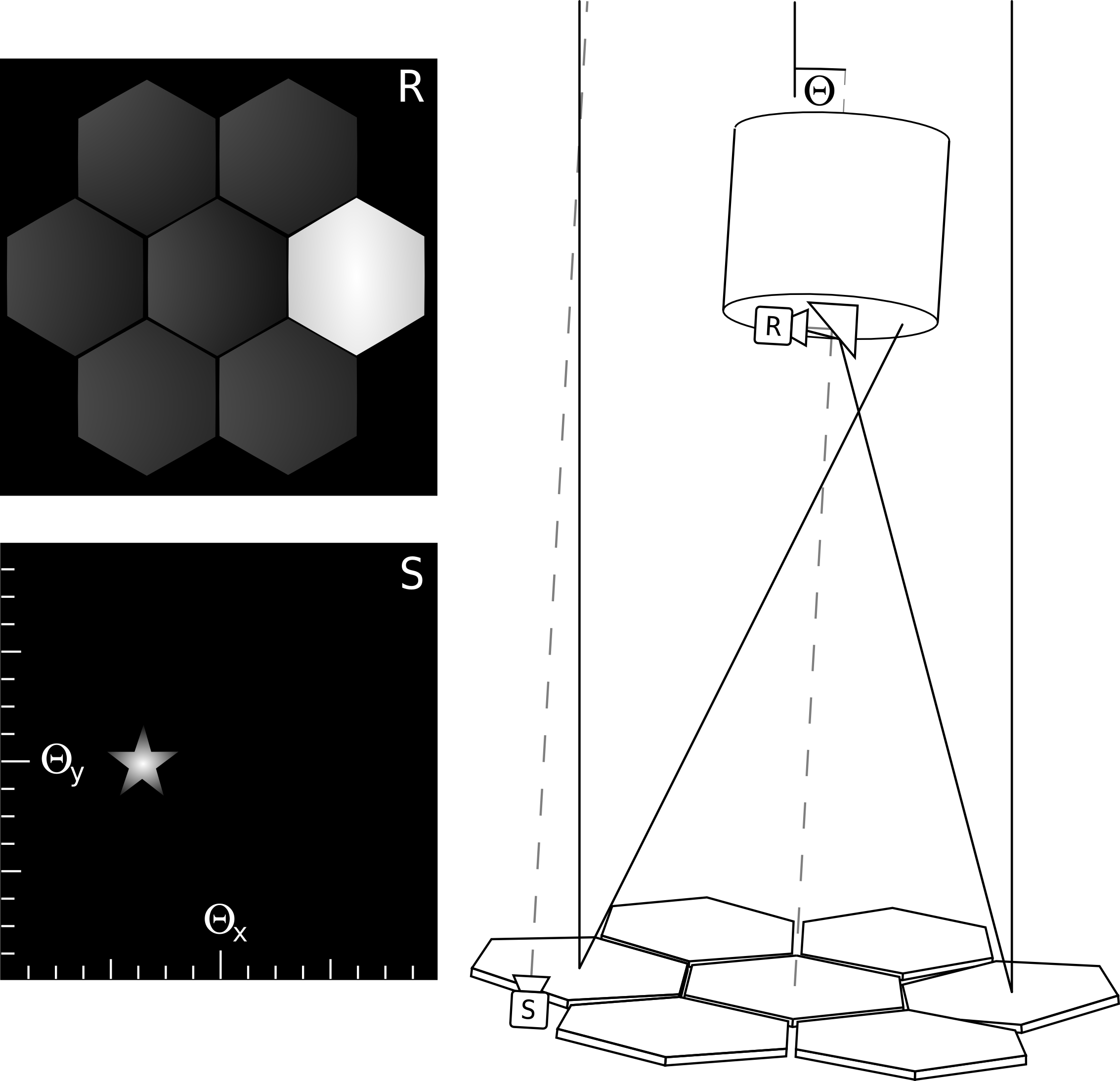}{%
    Light rays and camera images when pointing off axis to a star. 
    Here the telescope tilt $\RelativePointing$ happens to make the misaligned facet reflect the star light into the reflector's focal point. 
    In the reflector camera image the misaligned facet shows up bright whereas the rest of the facets are dark. 
    In the star camera image, the star appears off center by the angel $\RelativePointing$.
}{FoffAxisNamodSchematic}
Our \acs{namod} computer program reads out and adjusts both cameras.
To determine the facet orientations, the telescope is pointed to a bright reference star in the night sky, see Figures \ref{FonAxisNamodSchematic} and \ref{FoffAxisNamodSchematic}.
While the telescope moves close to the star, our \acs{namod} program, asynchronously from the telescope's drive, acquires images of the two cameras.
Both random or spiral telescope movements close to the star are possible.
The size of the region to move in depends on the pre-alignment state of the reflector. 
The better the pre-alignment, the smaller is the region to move in, the faster the \acs{namod} acquisition is done, see Section \ref{SecFactAlignmentResult}.
To later reconstruct the \acsp{psf} of the mirror facets, our \acs{namod} program records the telescope's relative pointing ${\RelativePointing_x}, {\RelativePointing_y}$ and the star's intensity $\StarIntensity$ as well as the mirror facets reflection intensities $\MirrorReflectionIntensity_{1}$\,...\,$\MirrorReflectionIntensity_{N}$ of all the $N$ facets.
Instead of communicating with the drive of the telescope, our \acs{namod} program restores ${\RelativePointing_x}, {\RelativePointing_y}$ directly from the star camera images.
\FigCapLab{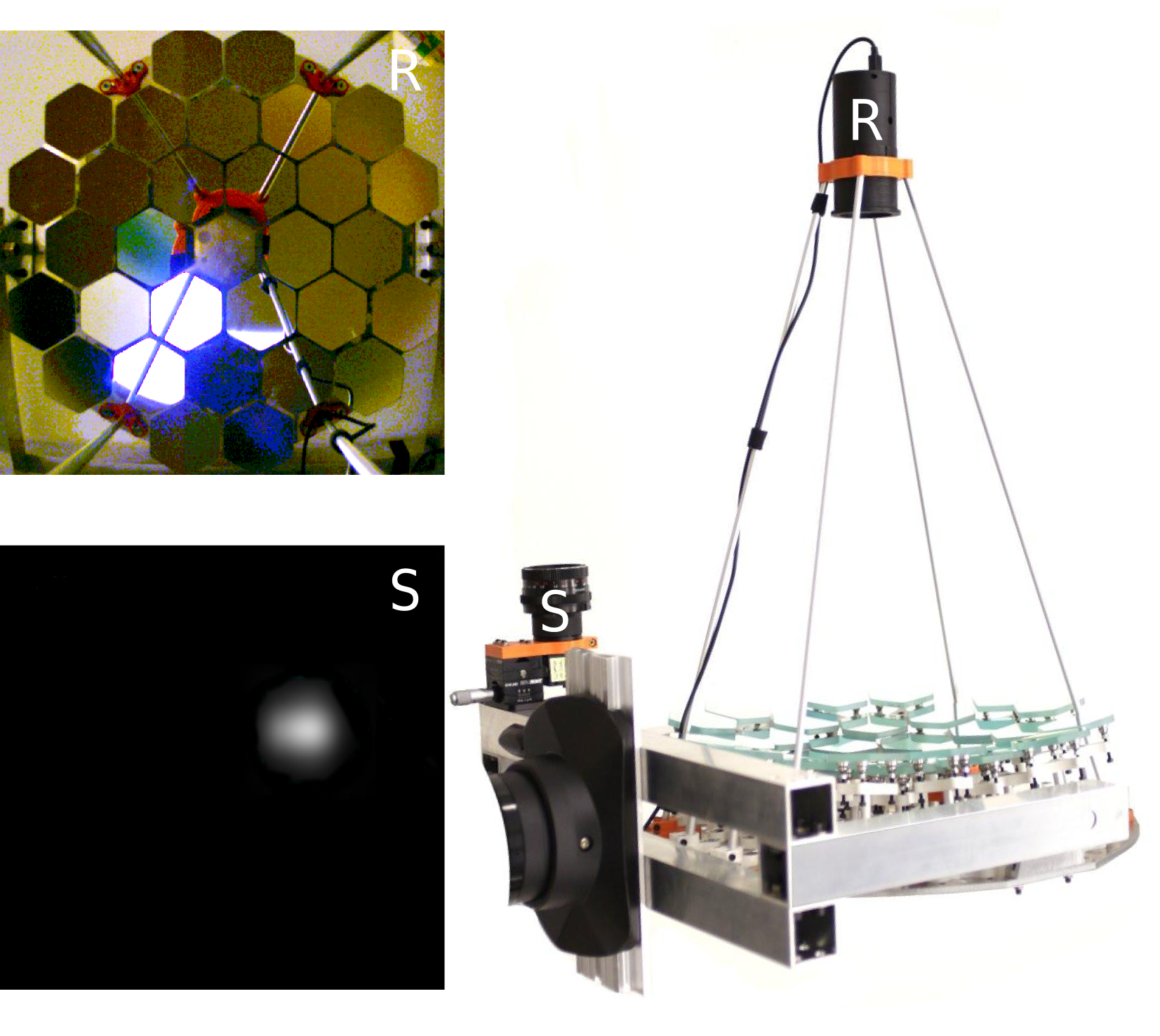}{%
    Actual reflector [R] and star [S] camera images of our \acs{namod} implementation running on the Mini \acs{fact} mock-up, see Section \ref{SecTestBench}.
    The star camera image is magnified here.
    Compare with the simplified overview in Figures \ref{FonAxisNamodSchematic} and \ref{FoffAxisNamodSchematic}.
}{f_overview_mini_fact}
During a \acs{namod} recording, the reference star's light intensity $\StarIntensity$ changes, e.g. due to clouds or varying zenith distance of the reference star. 
Also, $\StarIntensity$ changes when switching to another star.
All these changes in $\StarIntensity$ reduce the capability of reconstructing the facet orientations because the light intensities observed in the reflections of the mirror facets $\MirrorReflectionIntensity$ depend on the reference star.
The \acs{namod} method addresses this challenge by normalizing $\MirrorReflectionIntensity$ using $\StarIntensity$, see Section \ref{SecNormalizingMirrorResponse}.
This way, the mirror facet orientations are recorded more independent of the sky quality or the reference star.   
After the recording is done: First the mirror facets \acp{psf} are reconstructed, see Section \ref{SecPointSpreadReconstruction}.
Second, the orientation of each facet is calculated using the reflection law and the reconstructed facet's \acsp{psf}, see Section \ref{SecCorrectionImplementation}.
An overview of our implementation and example images are shown in Figure \ref{f_overview_mini_fact}.
%
\subsection{Cameras, calibration and control}
\label{SecCameraCalibration}
We use two industrial digital cameras for our \acs{namod} implementation, see Table \ref{Tcameras}.
First, a highly responsive $0.7\,$Mpixel camera and a wide angle lens serve as reflector camera, see Figure \ref{FReflectorCamOnMiniFACT}.
Second, a $5\,$Mpixel camera with a high mapping quality lens is used as star camera, see Figure \ref{FstarCamOnMiniFACT}.
We use the same cameras on Mini \acs{fact} and \acs{fact}.
\SideBySide{%
    \FigCapLab{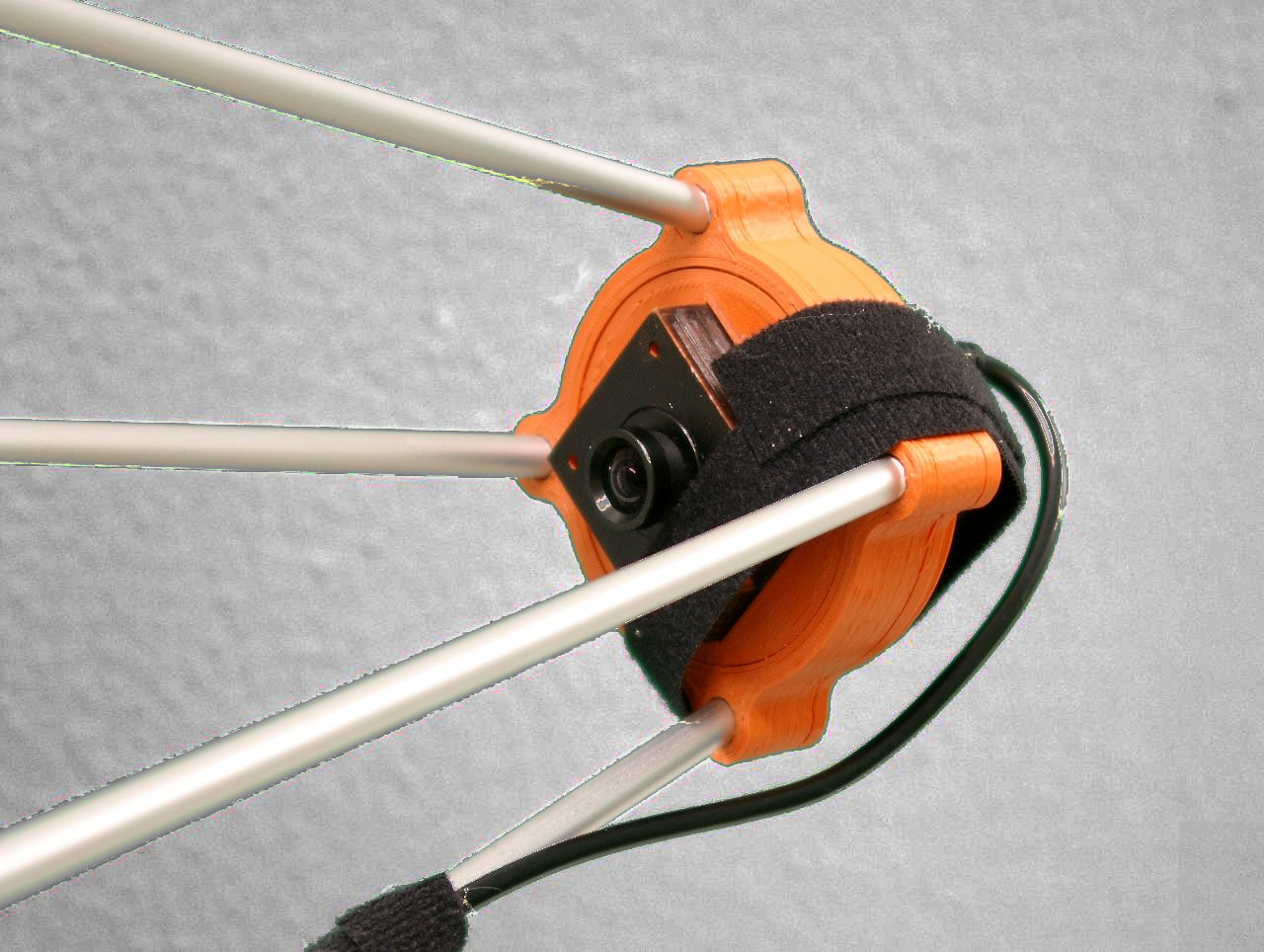}{%
        Reflector camera mounted on Mini \acs{fact} directly in the focal point.
    }{FReflectorCamOnMiniFACT}
}{%
    \FigCapLab{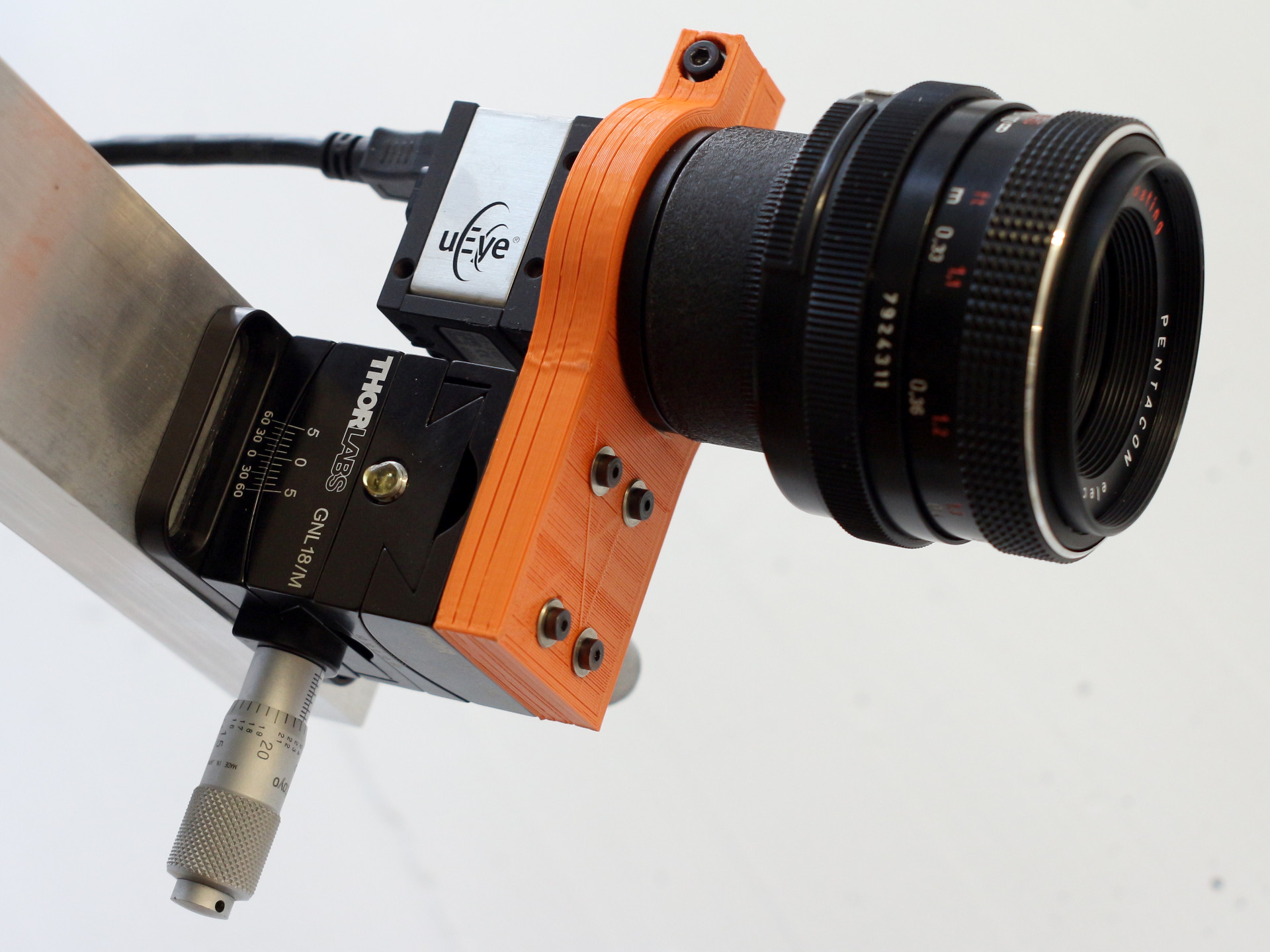}{%
        Star camera mounted on Mini \acs{fact} using a fine pan and tilt joint.
    }{FstarCamOnMiniFACT}
}
\begin{singlespace}
    \begin{table}[H]
    	\begin{tabular}{lrr}
             & reflector camera & star camera\\
            \toprule
            Type designation & IDS DCU223C & IDS UI-1480SE-M-HQ\\
            Optics $F\,[1]/f\,$[mm] & Im. Source $2.0/4.3$  & Zeiss Pentacon $1.8/50$\\
            Sensor/color & CCD/RGB & CMOS/B\&W \\
            Resolution [pixels] & $1024 \times 768$ & $2560 \times 1920$\\
            Communication & USB 2.0 & USB 2.0 \\
            FoV short edge [Deg] & $46.9$ & $4.34$\\
            Radiom. $\RadioCoeffCresponse$,\,$\RadioCoeffTexp$,\,$\RadioConstant$ & $1.03$,$\,-1.15$,$\,-0.19$ & $1.20$,$\,-1.23$,$\,0.99$\\
            pixel resolution $\geom\,[$mDeg/pixel$]$ & - & $2.437 \pm 0.001$\\
            \bottomrule
        \end{tabular}
        \caption[]{
    		Technical specifications for our reflector and star camera.	
        }
        \label{Tcameras}
    \end{table}
\end{singlespace}
To compare the light intensity observed in both cameras, the radiometric properties of the cameras have been measured.
Knowing the exposure time $\texp$ and the response of a pixel $\cresponse$, our \acs{namod} program calculates the absolute light intensity $\iflux$,  which is proportional to the photon intensity in this pixel.
In our implementation we use
\begin{eqnarray}
    \label{EqRadiometric}
    \log_{10}(\iflux) &=& \RadioCoeffCresponse \cdot \log_{10}(\cresponse) +    \RadioCoeffTexp \cdot \log_{10}(\texp) + \RadioConstant
\end{eqnarray}
to reconstruct $\iflux$.
The two slopes $\RadioCoeffCresponse$ and $\RadioCoeffTexp$ as well as the offset $\RadioConstant$ are measured in the lab by taking exposure time series while facing a reference light source at various distances with the bare image sensors.
More than three orders of magnitude in light intensity $\iflux$ are taken into account during calibration, and thus makes our camera and lens combinations sensitive to stars from about $-5$ down to $+3$ in apparent magnitude.
To estimate $\iflux$, the camera response $\cresponse$ must not overflow or underflow.
Our \acs{namod} program runs a feedback loop, which alters the exposure times $\texp$ of the star and reflector camera such that both maximum responses $\cresponse$ are always close to saturation.
Over or under saturated images are neglected automatically followed by an exposure time adjustment and reacquisition.
Typical acquisition rates for our \acs{namod} implementation are $0.2\,$Hz to $1\,$Hz, depending on the star and the sky quality, while $\texp$ is about $1\,$ms up to $3\,$s.\\
A calibration of the geometric properties of the star camera together with its lens is taken into account in order to later reconstruct the relative pointing directions $\RelativePointing_x$, $\RelativePointing_y$ from the star camera images.
Since the star camera lens distortion turned out to be negligible, an affine mapping relation is used where the angular resolution of a pixel $\geom$ is shown in Table \ref{Tcameras}.
For radiometric and geometric calibration we use methods inspired by \cite{Computer_Vision_Book_Forsyth_Ponce} and \cite{3D_Computer_Vision_Woehler_2012}.
%
\subsection{Reference star identification}
\label{SecStarRecognition}
Our \acs{namod} program identifies stars in the star camera image to estimate the relative telescope orientation $\Theta_x$, $\Theta_y$ and to measure the star's intensity $\StarIntensity$.
Our program accepts star images where only a single dominant star is present and where the shape and the size of the star is within an expected margin.
In our implementation we look for dominant stars with intensities $5$ standard deviations above the initial image noise level and accept spatial sizes of about $0.025^\circ$.
If a star image is rejected, both star and reflector image are discarded and acquired again.
%
\subsection{Mirror facet identification}
\label{SecFeedingFacets}
The \acs{namod} method needs to know which pixel intensities ${\iflux}_i$ of the \mbox{$i$-th} reflector image belong to a specific mirror facet $j$ so that the mirror facets reflection intensity $\MirrorReflectionIntensity_{i,j}$ can be averaged from these pixel intensities.
In our implementation, the \acs{namod} program is fed polygons describing mirror facets.
The polygons can have individual shapes and sizes.
To define the polygons, a reflector camera image is used while all the facets edges are visible, see Figure \ref{fInstructionGraphic}.
%
\subsection{Normalizing the mirror responses}
\label{SecNormalizingMirrorResponse}
In our \acs{namod} implementation, we extract for each record $i$ the reflection intensity $\MirrorReflectionIntensity_{i,j}$ of the \mbox{$j$-th} mirror facet from the reflector camera image as well as the star's intensity $\StarIntensity_i$ from the star camera image.
We obtain the normalized mirror facet reflection response
\begin{eqnarray}
    \label{EqNorm}
    \NormalizedMirrorResponse_{i,j} &=& \frac{m_{i,j}}{s_i}
\end{eqnarray}
by division.
Figure \ref{fIntensityNormalization} shows the performance of our radiometric camera calibration and the resulting normalization stability $\NormalizedMirrorResponse(\StarIntensity)$.
%
\subsection{\acf{psf} reconstruction}
\label{SecPointSpreadReconstruction}
\newcommand{\Hraw}{H_\text{raw}}
\newcommand{\Hexp}{H_\text{exposure}}
\newcommand{\Hpsf}{H_\text{\acs{psf}}}
By recording the normalized reflected intensity $\NormalizedMirrorResponse_j$ of the \mbox{$j$-th} mirror facet for many different pointings $\RelativePointing_x$ and $\RelativePointing_y$, the \acs{namod} method records the \acs{psf} of the \mbox{$j$-th} mirror facet directly.\\
In our implementation of \acs{namod}, we export the mirror facets \acsp{psf} using $2$D histograms $\Hpsf$ because these are easy to interpret and work with.
To fill the final $\Hpsf$, we take the exposure map of the pointings $\RelativePointing_x$ and $\RelativePointing_y$ into account.
This is done as the telescope movement might be unevenly spread along the $2$D pointing range.
First, we fill for each mirror facet $j$ a weighted $2$D histogram $\Hraw^j$.
The normalized mirror response $R_{i,j}$ of each record $i$ is filled into the bins of $\Hraw^j$ according to the corresponding pointing ${\RelativePointing_x}_i$ and ${\RelativePointing_y}_i$.
Second, we fill a $2$D exposure histogram $\Hexp$ to count the numbers of records taken in a specific pointing bin.
We then obtain the final $2$D \ac{psf} histogram $\Hpsf^j$ for each mirror facet $j$, by bin wise dividing the facet's raw response using the exposure histogram: $\Hpsf^j = \Hraw^j/\Hexp$.
Figure \ref{FigPsfs14vs15} shows example \acs{psf} histograms recorded with our \acs{namod} implementation.
%
\subsection{Correction Implementation}
\label{SecCorrectionImplementation}
\newcommand{\ThetaCoG}[1]{\Theta_{\tiny{\acs{cog}},#1}}
\newcommand{\PhiCoG}[1]{\varPhi_{\tiny{\acs{cog}},#1}}
After recording and reconstruction of the individual mirror facet \acsp{psf} $\Hpsf$, our \acs{namod} implementation produces both human and machine readable instructions to correct the facet misalignments.
First, we calculate the position $\ThetaCoG{j}$ of the \ac{cog} in  $\Hpsf^j$ for each facet $j$ with respect to the reflectors focal point.
Second, we calculate the correction angle to be applied to the \mbox{$j$-th} mirror facet $\PhiCoG{j}$ using the reflection law
\begin{eqnarray}
    \label{EqCorrection}
    \PhiCoG{j} &=& - \frac{1}{2} \ThetaCoG{j}.
\end{eqnarray}
Our \acs{namod} program knows the inverse kinematics of the mirror facet mountings of \acs{fact} so it can further give directly the manipulation instructions for the three linear joints of a mirror facet's tripod mount.
Figure \ref{fInstructionGraphic} shows an example of our \acs{namod} program's instructions.
%
\subsection{The \acs{namod} development mock-up -- Mini \acs{fact}}
\label{SecTestBench}
Our \acs{namod} implementation was developed on a mock-up called Mini \acs{fact} in order to not lose observation time on \acs{fact}.
Mini \acs{fact} is a 10th scale model of FACT with similar tripod mirror facet mounting, see Figure \ref{FminiFACT} and Table \ref{t_mini_fact_optical_data}.
It has a fully operational segmented reflector of 30 facets and has a angular resolution comparable to \acs{fact}.
\begin{singlespace}
    \begin{table}[H]
        \begin{tabular}{lr}
            \toprule
            Focal length and F-number & $450\,$mm, $1.29$\\
            Aperture & $30$ hexagons, each $31.7\,$cm$^2$, total $951\,$cm$^2$\\
            Geometry & Davies-Cotton and/or Paraboloid\\
            Mirror facets & Float glass, front aluminum coated, spherical\\
            \bottomrule
        \end{tabular}
        \caption[]{
            Optical properties of Mini \acs{fact}
        }
        \label{t_mini_fact_optical_data}
    \end{table}
\end{singlespace}
\FigCapLab{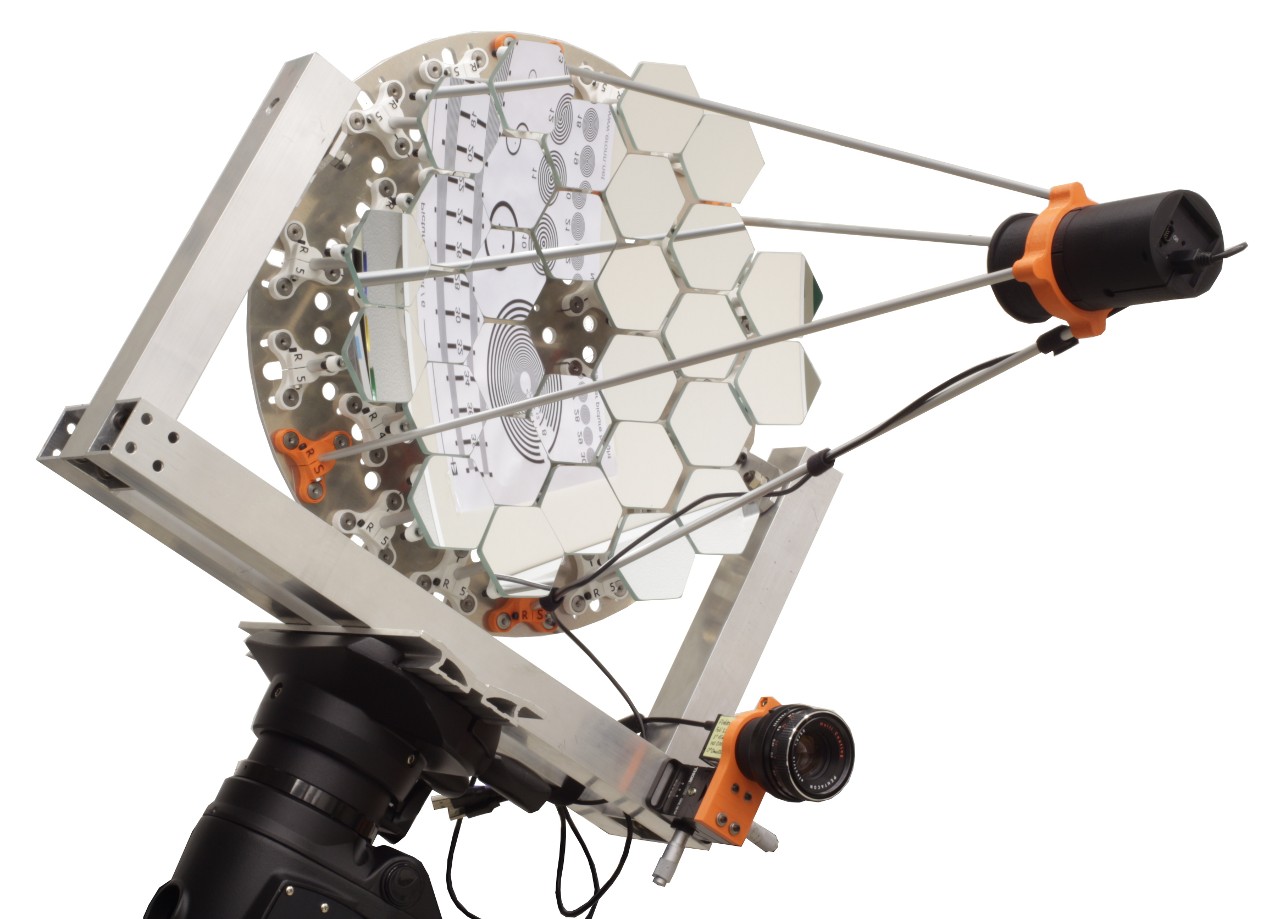}{
    The 10th scale model Mini \acs{fact}.
    All our \acs{namod} components can be mounted to it and were tested in the lab and under the night sky.
}{FminiFACT}
%
\section{Results}
\label{SecFactAlignmentResult}
%
\subsection{Fine alignment of \acs{fact} in May 2014}
\FigCapLab{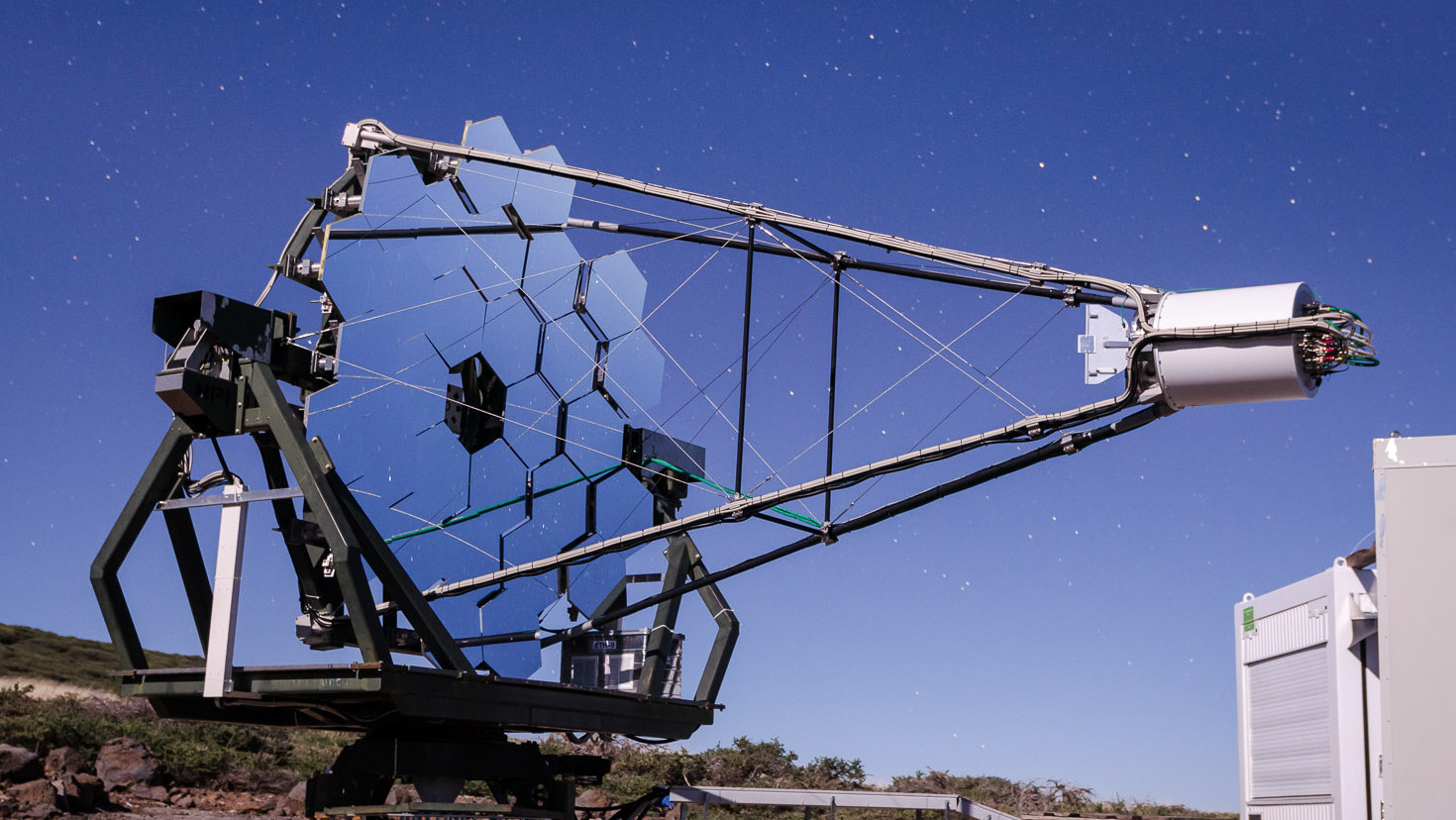}{
    \acs{fact} is located on the Canary island La Palma, Spain.
    It inherited its mount and the mirror facets from HEGRA \cite{HEGRA_status_and_results_1998}. 
    While pioneering silicon photomultipliers for \acs{iact}s, \acs{fact} is monitoring gamma ray bright Blazars such as Mrk\,421 and Mrk\,501. 
    Photograph by Thomas Kr\"ahenb\"uhl.
}{fFACT}
\begin{table}[H]
    \begin{center}
        \begin{tabular}{lr}
            \toprule
                focal length $f$ & $4.889\,$m\\
                number of facets & $30$\\
                facet mounting & manual adjustment on tripod\\
                %
                reflector geometry along optical axis & $1/2$\,Davies Cotton + $1/2$\,parabola\\
                reflector area $A$ & $9.51\,$m$^2$\\
                effective reflector area $A_\text{eff}$ & $8.80\,$m$^2$\\
                effective aperture diameter $D_\text{eff}$ & $3.35\,$m\\
                maximum aperture diameter $D_\text{max}$ & $3.93\,$m\\
                effective F-number, $f/D_\text{eff}$ & $1.46$\\
                F-number, $f/D_\text{max}$ & $1.25$\\
                image sensor diameter, FoV & $0.39\,$m, $4.5^\circ\,$Deg\\
            \bottomrule
        \end{tabular}
        \caption[]{Basic imaging reflector properties of \ac{fact}}
        \label{TabFact}
    \end{center}
\end{table}
\subsubsection*{Reflector redesign and need for new alignment}
In May 2014 we redesigned the \acs{fact} reflector, see Figure \ref{fFACT} and Table \ref{TabFact}, to be a hybrid of Davies Cotton \cite{Davies_Cotton_1957} and paraboloid geometry to decrease the reflector's time spread with an acceptable worsening of its spatial resolution.
Since \acs{fact}'s overall timing precision of $\approx 0.3\,$ns used to be in the regime of the reflector's time spread, we changed the reflector's geometry to further improve the overall timing precision \cite{FACT_Muon_calibration_ICRC2015} and lower the minimal energy trigger threshold.
\subsubsection*{Numerical \acs{psf} comparison}
For numerical guidance, we compare the areas $A_{\sigma}$ of the different \acsp{psf} before and after the \acs{namod} alignment, see Table \ref{TabSigmaArea}.
The area
\begin{eqnarray}
    A_{\sigma} &=&  \pi \sigma_a \sigma_b
    \label{Asigma}
\end{eqnarray}
is defined by the ellipse spanned by the standard deviations
\begin{eqnarray}
    \sigma_a, \sigma_b &=& \sqrt{ \text{eig}\left( \text{cov}(\iflux(\RelativePointing_x, \RelativePointing_y)) \right) }
    \label{psf_eig}
\end{eqnarray}
of the light intensity distribution $\iflux(\RelativePointing_x, \RelativePointing_y)$ along its principal components.
Here $\text{eig}(M)$ gives the eigenvalues of matrix $M$ and $\text{cov}(I)$ gives the covariance matrix of the distribution $I$.
The ellipses and the corresponding principal component directions are highlighted in red in the Figures \ref{FpsfAfterBokeh2014}, \ref{FpsfAfterSCCAN2014} and \ref{FpsfRayTracing}.
\subsubsection*{pre-alignment}
After the redesign, we performed a first pre-alignment using the most simple version of Bokeh alignment \cite{FACT_Bokeh_alignment}, which led to the overall \acs{psf} shown in Figure \ref{FpsfAfterBokeh2014}.
The small pre-alignment \acs{psf} sped up the following \acs{namod} alignment since the movements close to the reference star could be limited to a region of the size of the pre-alignment \acs{psf} of $\approx 0.5^\circ$ in diameter. 
\subsubsection*{\acs{namod} alignment}
After pre-alignment, our \acs{namod} alignment fine tuned the \acs{fact} reflector in a single iteration resulting in the \acs{psf} shown in Figure \ref{FpsfAfterSCCAN2014}.
For this particular \acs{namod} alignment, our implementation took about 1300 records in $1\,$hour.
Figure \ref{fInstructionGraphic} shows the instructions provided by our \acs{namod} implementation, which were applied to \acs{fact}'s mirror tripod mount joints manually using a goniometer.
Deformation tests on FACT showed, that gravitational slump is not an issue on its reflector, but to be sure in this first \acs{namod} alignment, we chose only reference stars within $45^\circ$ zenith distance.
\subsubsection*{Theoretical lower \acs{psf} limit}
For comparison we present the lower limit of the \acs{fact} reflector \acs{psf} found in ray tracing simulations in Figure \ref{FpsfRayTracing}. 
In the simulation, the \acs{fact} reflector has perfect spherical mirrors, a perfect alignment and the actual hybrid Davies Cotton and parabola geometry.
\subsubsection*{Direct \acs{psf} recording with dedicated image sensor}
The \acsp{psf} in figures \ref{FpsfAfterBokeh2014} and \ref{FpsfAfterSCCAN2014} are recorded with our radiometrically calibrated \mbox{$6 \times 6\,$cm$^2$} digital image sensor, that is placed in \acs{fact}'s pseudo focal plane while tracking the star Arcturus.
This image sensor is made out of a vintage medium format camera's view finder screen (Hasselblad $6 \times 6$), which is observed by an industrial CCD camera and has an effective resolution of \mbox{$667 \times 667$\,pixels}, respectively \mbox{$1.05\,$mDeg/pixel} when mounted on the \acs{fact} reflector.
\begin{table}[H]
    \begin{center}
        \begin{tabular}{lrr}
                Reflector state &  $A_{\sigma}$ [arcmin$^2$] & relative [\%]\\
            \toprule
                before reconfiguration & $62.0$ & 1088\\
                after reconfiguration & too large to be recorded & too large\\
                after Bokeh alignment & $65.5$ & 1149\\
                after \acs{namod} alignment & $14.8$ & 260\\
            \midrule
                ray tracing, perfect reflector & $5.7$ & 100\\
            \bottomrule
        \end{tabular}
        \caption[]{
            The \acs{fact} on axis \acsp{psf}, see Figures \ref{FpsfAfterBokeh2014}, \ref{FpsfAfterSCCAN2014} and \ref{FpsfRayTracing}.
        }
        \label{TabSigmaArea}
    \end{center}
\end{table}
\FigCapLab{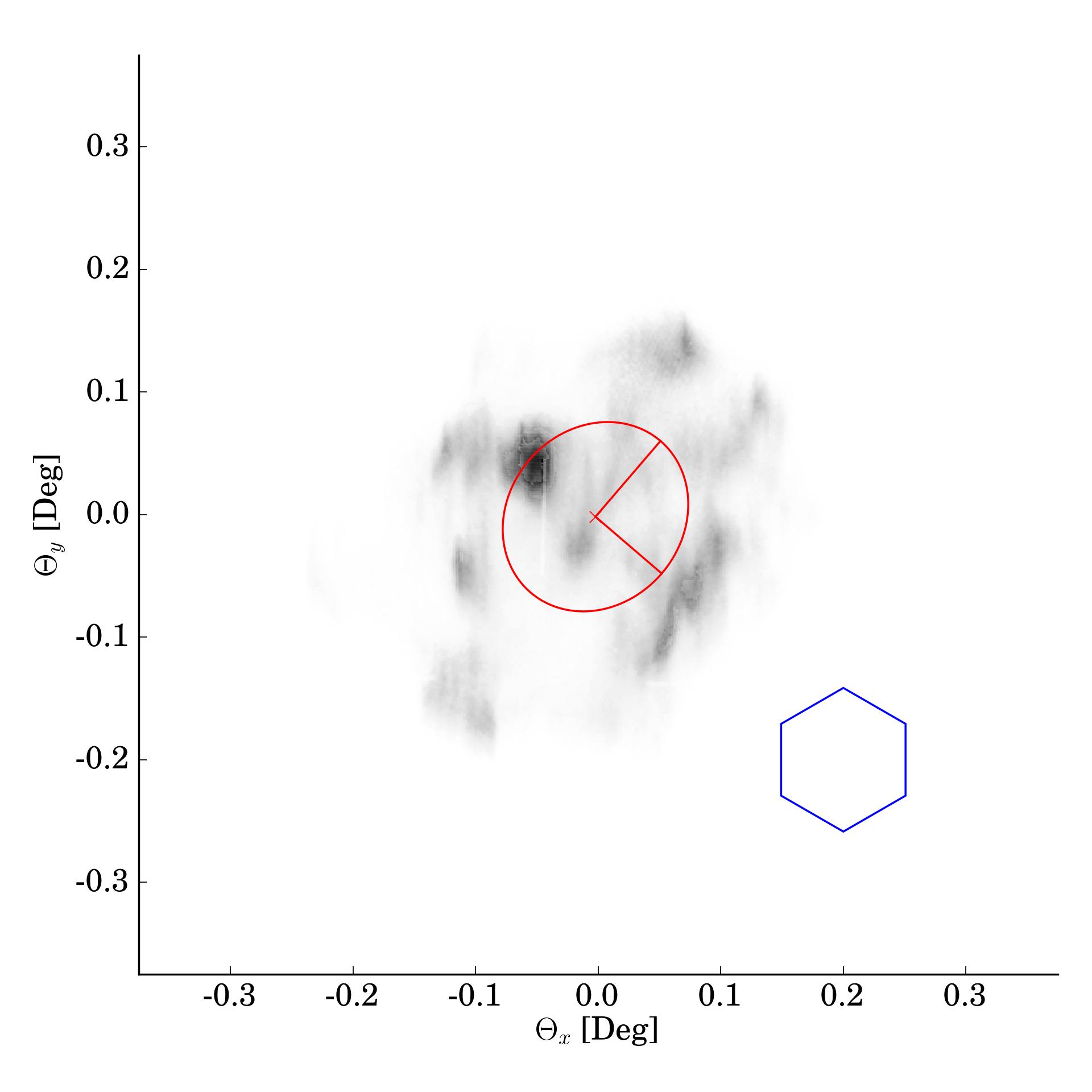}{
    \acs{fact} \acs{psf} after Bokeh pre-alignment. Overlaid with confinement ellipsis \mbox{$A_{\sigma} = 65.5$\,arcmin$^2$}. The hexagon represents the aperture of a \acs{fact} pixel.}
{FpsfAfterBokeh2014}
\FigCapLab{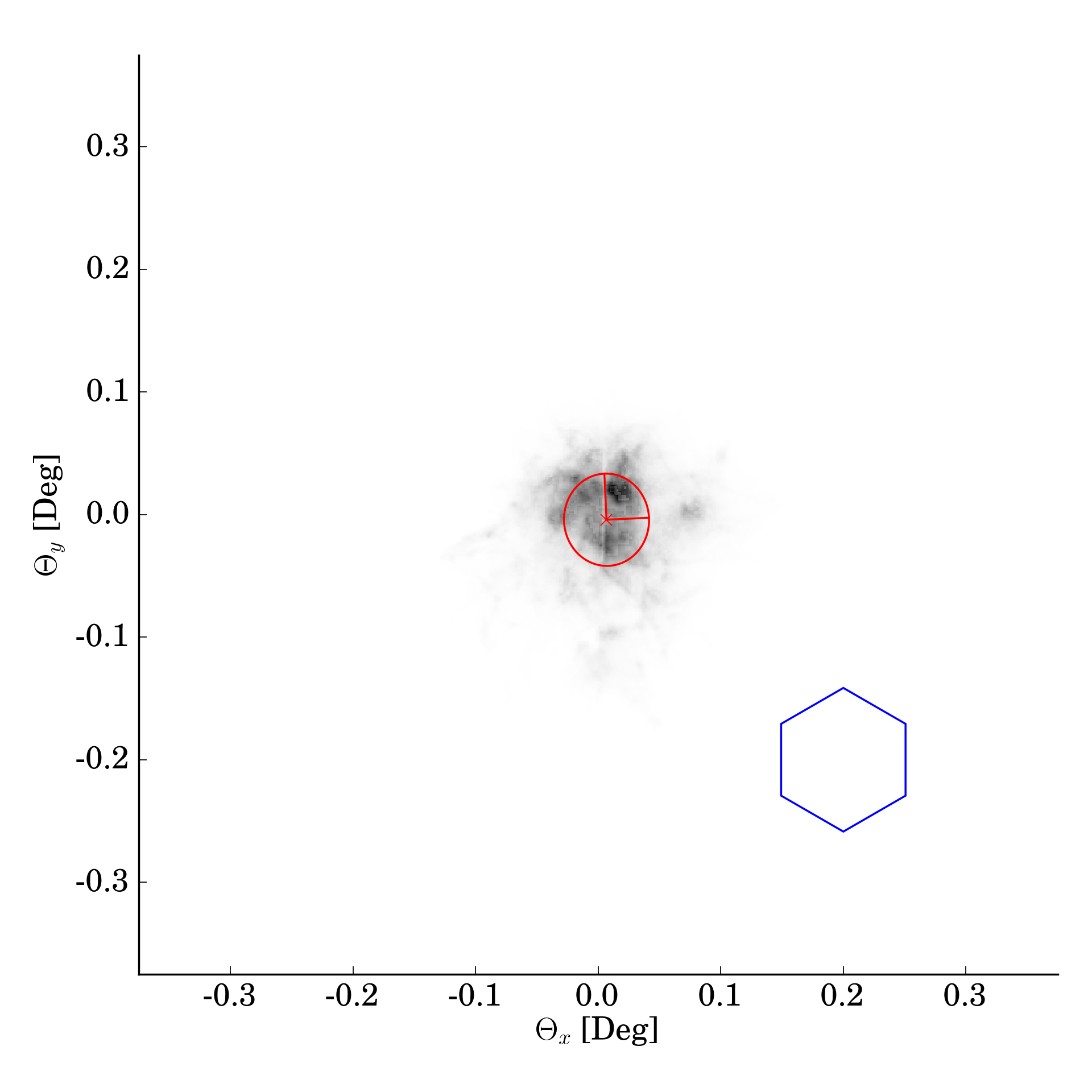}{
    \acs{fact} \acs{psf} after \acs{namod} alignment. \mbox{$A_{\sigma} = 14.8$\,arcmin$^2$}}
{FpsfAfterSCCAN2014}
\FigCapLab{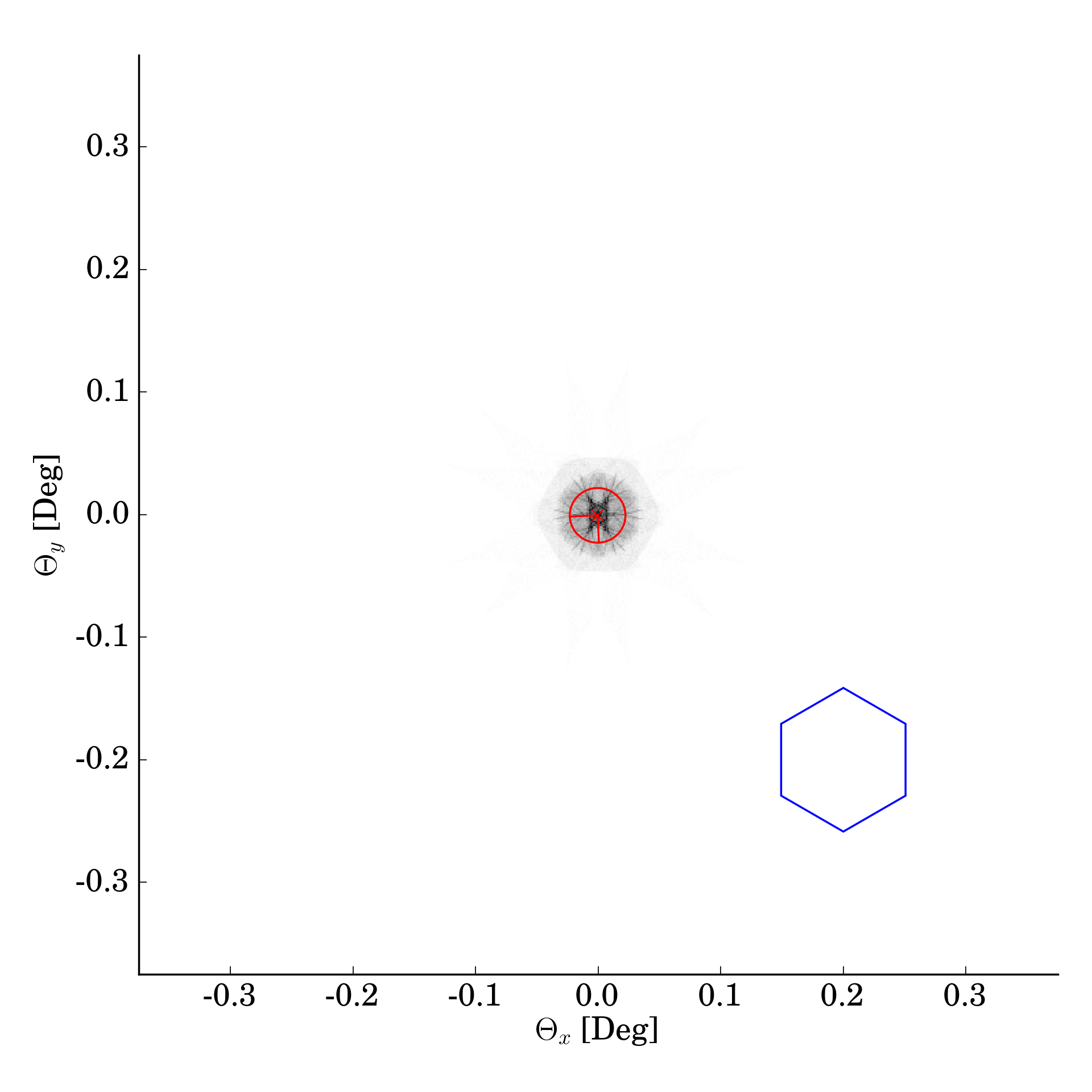}{
    Ray tracing simulation of the optimum \acs{fact} \acs{psf}. \mbox{$A_{\sigma} = 5.7$\,arcmin$^2$}}
{FpsfRayTracing}
\FigCapLab{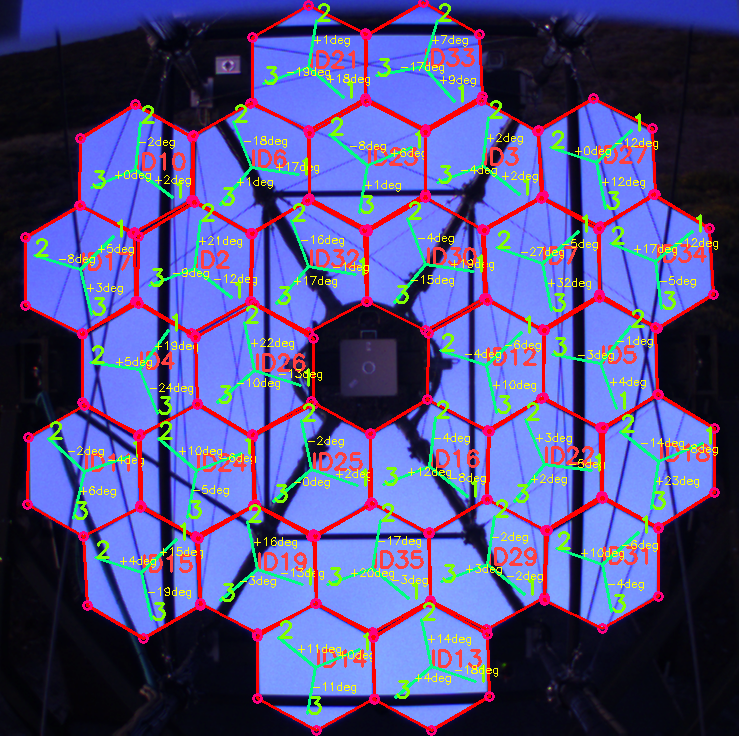}{
    Reflector image with instructions to improve alignment, created by our \acs{namod} implementation.
    Mirror IDs are shown in the center of the mirror facets boundary polygons, both are shown in red.  
    Green mirror tripod mount and arm number show the actual orientation.
    Close to a tripod arm, the correction turn angle for the threaded bolt of the linear joint is given in yellow.
}{fInstructionGraphic}
%
\subsection{Independence of night sky conditions}
Two years after the initial radiometric calibration of the two cameras in our \acs{namod} implementation, we checked the normalization again, in the lab, and directly on a Mini \acs{fact} facet.
Mini \acs{fact} was pointing to an artificial light source of adjustable intensity while our \acs{namod} setup was taking records.
Figure \ref{fIntensityNormalization} shows the recorded 'star' intensity $\StarIntensity$ and the normalized mirror intensity $\NormalizedMirrorResponse(\StarIntensity)$. 
The normalized mirror intensity $\NormalizedMirrorResponse$ changes only $(17.2\pm0.5)\%$ when $\StarIntensity$ changes by one order of magnitude, respectively it changes $(6.9\pm0.2)\%$ for one step of $\StarIntensity$ in apparent magnitude.
\FigCapLab{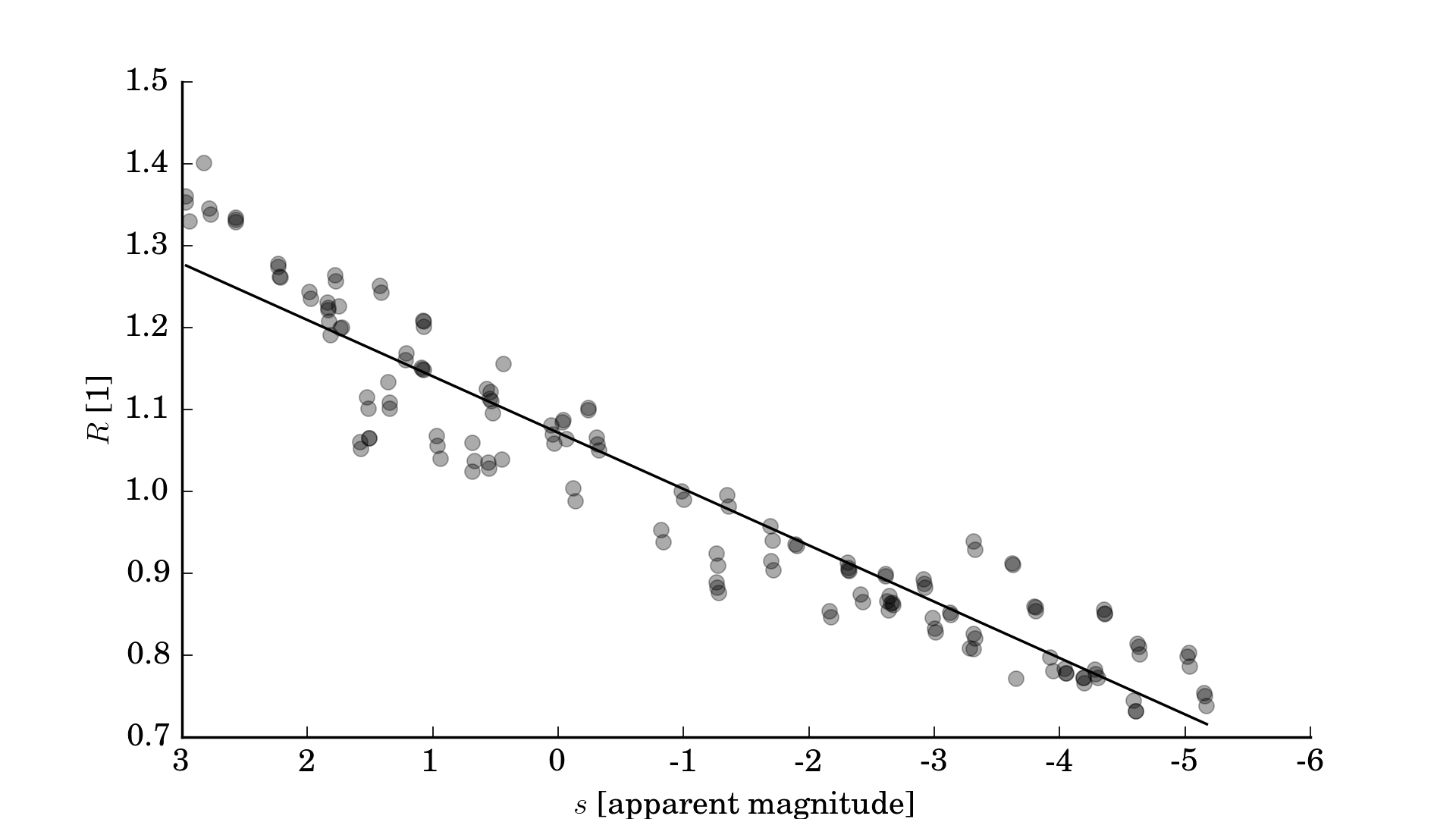}{
    Measured normalization stability of our \acs{namod} implementation. The black line shows the fitted change in normalized response $\NormalizedMirrorResponse$, which is $(-6.9\pm0.2)\%$ for a brightening of $\StarIntensity$ by one apparent magnitude. Ideally $\NormalizedMirrorResponse(\StarIntensity)$ should be flat.
}{fIntensityNormalization}
Our \acs{namod} implementation aligned Mini FACT several times successfully on the partly clouded night sky of Dortmund, Germany.
Unusable records were rejected automatically and we did not notice a drop in \acs{psf} reconstruction quality although the reference stars and their intensities changed during the process. 
This shows, that our \acs{namod} implementation's intensity normalization helps to reconstruct the facets orientations more independent of the sky conditions across the full range of possible reference star magnitudes.

\subsection{\ac{psf} reconstruction performance}
To demonstrate the performance and repeatability of our \acs{namod} implementation, we show that the individual mirror facet \acs{psf} signatures can be identified for two \acs{namod} runs, separated by one year, see Table \ref{t14vs15}.
Without moving or tilting the facets, our \acs{namod} implementation recorded the facets orientations to feed the \acs{fact} \acs{iact} simulation so that mismatches between observed and simulated air shower records can be reduced.
\begin{singlespace}
    \begin{table}[H]
        \begin{center}
            \begin{tabular}{lrr}
                 & May 2014 & May 2015\\
                \toprule
                records taken & 1\,k & 5\,k\\
                reference object & star Arcturus & planet Jupiter\\
                zenith distance & $\approx 20\,^\circ$  & $\approx 40\,^\circ$ \\
                recording time & $58\,$min & $2\,$h $18\,$min \\
                \bottomrule
            \end{tabular}
            \caption[]{
                \acs{namod} runs not used for alignment but to feed the \acs{fact} telescope simulation.
            }
            \label{t14vs15}
        \end{center}
    \end{table}
\end{singlespace}
Figure \ref{FigPsfs14vs15} shows a sample of reconstructed $\Hpsf$s for individual \acs{fact} mirror facets.
The more dense $\Hexp$ of the year 2015 reveals the \acs{psf}s of this year in more detail.
\begin{figure}[H]
    \begin{center}
        \TwoFigsSideBySide{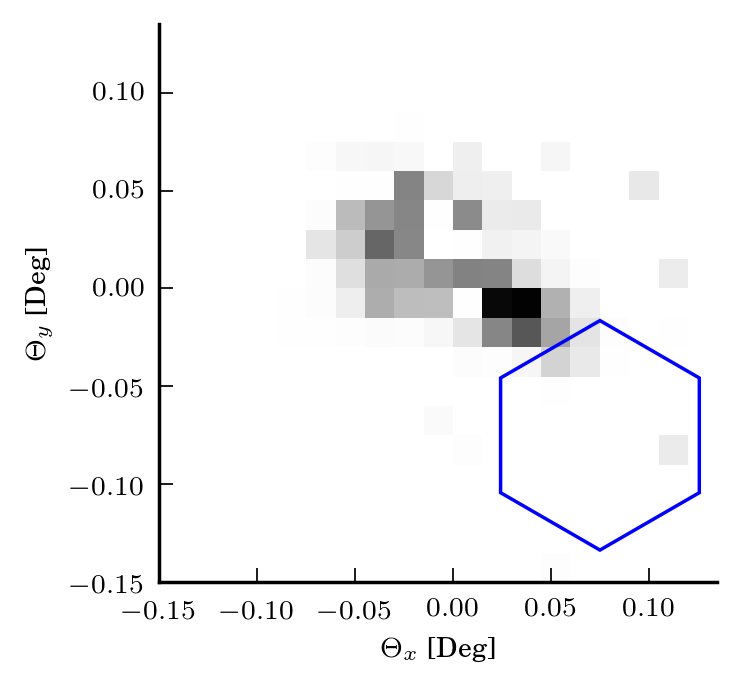}{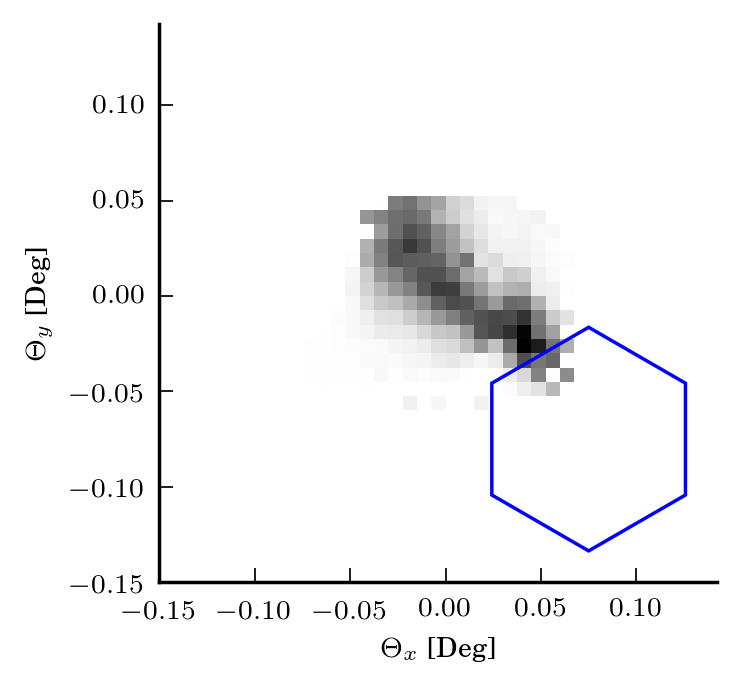}
        \TwoFigsSideBySide{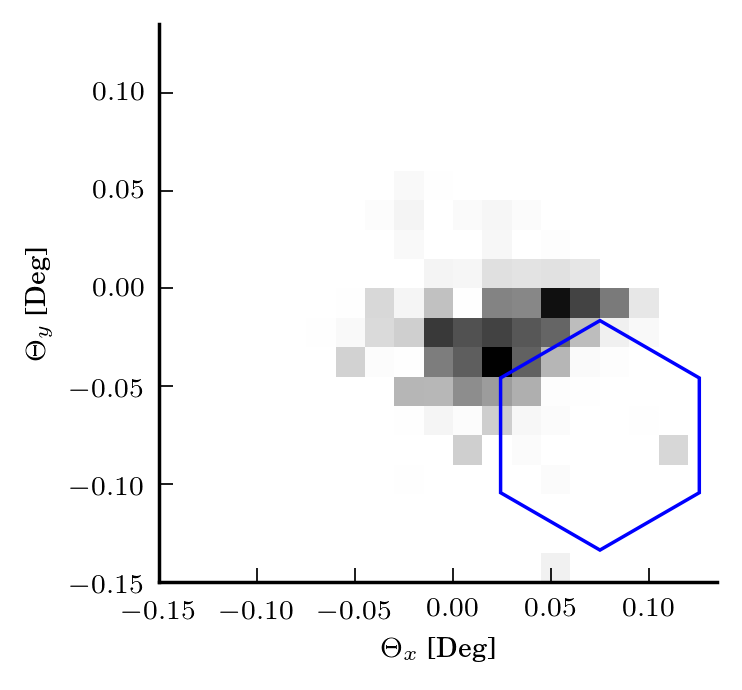}{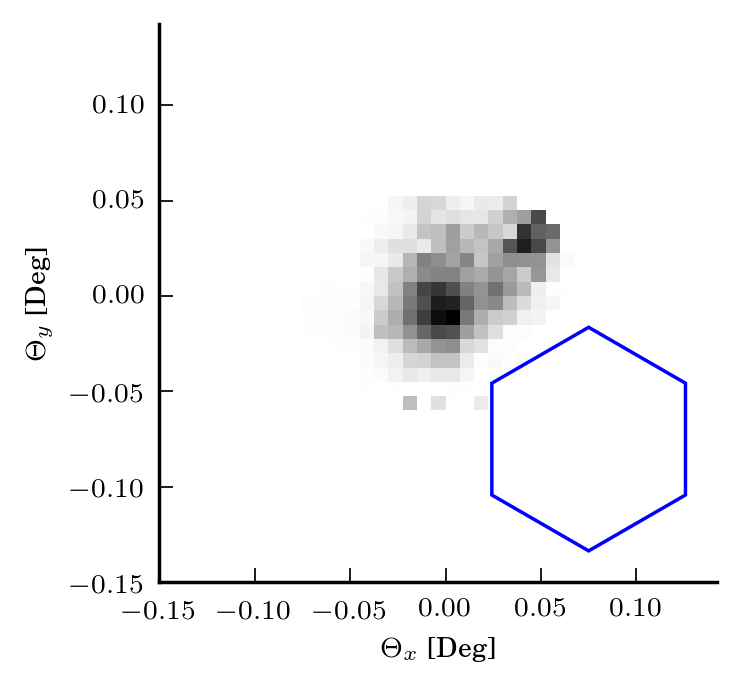}
        \TwoFigsSideBySide{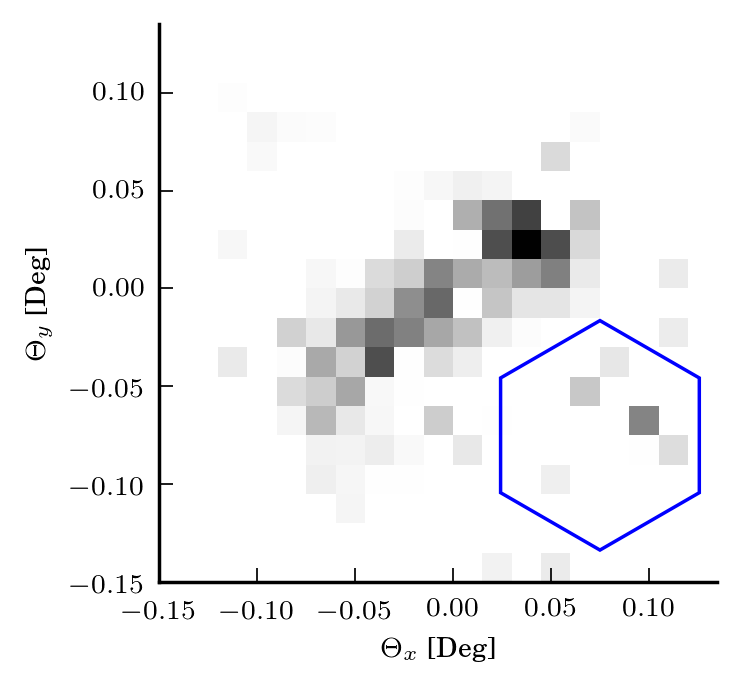}{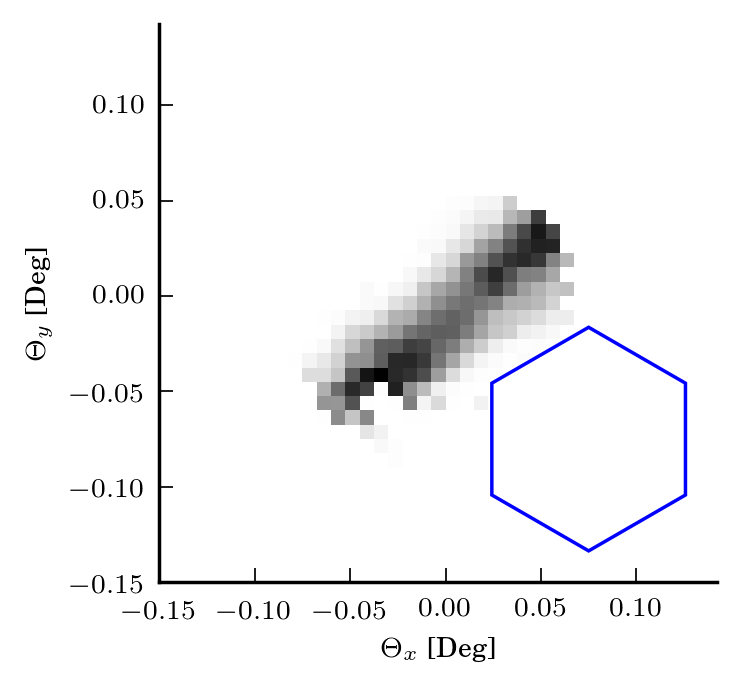}
        \caption[]{%
            $\Hpsf$ for individual \acs{fact} mirror facets.
            Left column: 2014, right column: 2015.
            Each row shows the same mirror facet for both dates.
            The fingerprints of the facets \acsp{psf} can be identified for both years.
        }
        \label{FigPsfs14vs15}
    \end{center}
\end{figure}
%
\section{Conclusion}
Our \acs{namod} implementation is stable and delivered a high quality \acs{psf} during its first use.
In a single iteration on \acs{fact}, our \acs{namod} implementation narrowed the \ac{psf} area down from $1149\%$ to only $260\%$ of the theoretical \ac{psf} area limit found in ray tracing simulations, where the ray tracing was done for perfect mirror facets and a perfect alignment.  
The mirror facet reflection normalization allows to switch to various reference stars during recording and makes facet orientation reconstruction less dependent on changes in sky quality or zenith distance.
The absence of telescope drive communication, and the flexible feeding of facet geometry makes our \acs{namod} implementation a simple to use, plug and play solution, that can be applied to similar telescopes easily.
For example it can be applied to Mini \acs{fact}, which runs a completely different drive software.
The pocket Mini \acs{fact} mock up sped up the development of our \acs{namod} implementation while no observation time was lost on \acs{fact}.
\newline
In future implementations one might install the \acs{namod} cameras permanently.
On \acs{fact}, our \acs{namod} implementation was mounted temporary but a permanent installation is possible when placing the reflector camera and the pseudo optical axis on top of the lid of the telescope's image sensor.
The reflection coefficients of the mirror facets might be determined as well.
The normalization stability $\NormalizedMirrorResponse(\StarIntensity)$ might be further improved by not using electric light bulbs but light sources with a spectral distribution closer to the one of stars.
Also, the \acs{namod} records might be used to validate the ray tracing \acs{iact} simulation when used as directional look up table for a given aperture intersection.
An adaptation of \acs{namod} on the dual mirror Schwarzschild Couder telescopes of the Cherenkov Telescope Array (CTA) \cite{CTA_Introduction} is not straight forward.
However, because of the \acs{psf} quality, but especially because of the flexibility and the tolerance for a wider range of night sky conditions, we believe \acs{namod} is ideal for the future single mirror (single primary segmented reflector) telescopes of CTA.
%
\section*{Acknowledgments}
The  important  contributions  from  ETH Zurich  grants  ETH-10.08-2  and  ETH-27.12-1  as  well  as  the funding by the German BMBF (Verbundforschung Astro- und Astroteilchenphysik) are gratefully acknowledged.
We are thankful for the very valuable contributions from E.~Lorenz, D.~Renker and G.~Viertel during the early phase of the project.
We thank the Instituto de Astrofisica de Canarias allowing us to operate the telescope at the Observatorio Roque de los Muchachos in La Palma, the Max-Planck-Institut fuer Physik for providing us with the mount of the former HEGRA CT 3 telescope, and the MAGIC collaboration for their support.
We would also like to thank the OpenCV community, the StackOverflow community, J. Hunter for Matplotlib, the GNU compiler community, IDS imaging for excellent product documentation and H. Mueller and J. Stratmann for discussions on vintage optics.
%
\section*{References}
\bibliography{references.bib}
%
\begin{acronym}
    \acro{cog}[CoG]{Center of Gravity}
	\acro{sccan}[SCCAN]{Solar Concentrator Characterization At Night}
	\acro{psf}[PSF]{Point Spread Function}
	\acro{fact}[FACT]{First Geiger-mode Avalanche Cherenkov Telescope}
	\acro{veritas}[VERITAS]{Very Energetic Radiation Imaging Telescope Array System}
    \acro{iact}[IACT]{Imaging Atmospheric Cherenkov Telescope}
    \acro{namod}[NAMOD]{Normalized and Asynchronous Mirror Orientation Determination}
\end{acronym}
\end{document}